# Hierarchical Grid-Based Pairwise Key Pre-distribution in Wireless Sensor Networks

Abedelaziz Mohaisen,* DaeHun Nyang†and KyungHee Lee‡

## Abstract

*The security of wireless sensor networks is an active topic of research where both symmetric and asymmetric key cryptography issues have been studied. Due to their computational feasibility on typical sensor nodes, symmetric key algorithms that use the same key to encrypt and decrypt messages have been intensively studied and perfectly deployed in such environment. Because of the wireless sensor's limited infrastructure, the bottleneck challenge for deploying these algorithms is the key distribution. For the same reason of resources restriction, key distribution mechanisms which are used in traditional wireless networks are not efficient for sensor networks.*

*To overcome the key distribution problem, several key pre-distribution algorithms and techniques that assign keys or keying material for the networks nodes in an offline phase have been introduced recently. In this paper, we introduce a supplemental distribution technique based on the communication pattern and deployment knowledge modeling. Our technique is based on the hierarchical grid deployment. For granting a proportional security level with number of dependent sensors, we use different polynomials in different orders with different weights. In seek of our proposed work's value, we provide a detailed analysis on the used resources, resulting security, resiliency, and connectivity compared with other related works.*

## 1 Introduction

Sensor network consists of a huge number of sensor nodes which are inexpensive, low-powered and resources-constrained small devices [5]. The typical sensor node contains a power unit, a sensing unit, a processing unit, a storage unit, and a wireless transceiver (T/R) [1]. The concept of micro-sensing and wireless connection in the sensor network promises several applications in military, environment, healthcare, and many other commercial domains [15]. Due to sensor nodes resources' constraints, public key algorithms such like Deffie-Hellman key agreement [7] or the RSA Signature [28] are undesirable to be used. In spite of recent results on the computational feasibility of those algorithms [11, 20, 32, 33], it is still early to widely deploy these algorithms since using them will expose a vulnerability to denial of service attack (DoS) [6, 34].

On the other hand, symmetric key algorithms that use same key for encrypting and decrypting messages are desirable in the sensor network. This desirability is due to the computational lightness on the typical sensors. From another point, due to the weak infrastructure of the sensor network, traditional secret key distribution mechanisms such like the Key Distribution Center (KDC) can not be used. The main issue therefore is summarized in how to distribute secret keys or keying material that are responsible on generating secret keys among different sensor nodes [10]. Since the manual modification of the sensors' contents is undesirable after the in-field deployment phase, several key *pre-distribution schemes* that assign and distribute keying material or secret keys in an off-line phase have been proposed. In the following section, we review some of those schemes followed by our main contribution.

### 1.1 Related Works

Two of the early works in [2, 3] are widely known for their novelty. Considering a network that consists

*Electronics and Telecommunications Research Institute, Daejeon 305-700, Korea

†Graduate School of Information Technology and Telecommunication, Inha University, Incheon 402-751, Korea, nyang@inha.ac.kr

‡Department of Electrical Engineering, The University of Suwon, Suwon, 445-743, Korea

of $N$ nodes, in the first work by Blom et. al. [2] a symmetric matrix of size $N \times N$ is required to store the different $N^2$ keys for securing communication within the entire network. Node $s_i \in N$ has row and column in the matrix. If two nodes $s_i, s_j$ would like to communicate, they use the entries $\mathbf{E_{ij}}$ in $s_i$ side and $\mathbf{E_{ji}}$ in $s_j$ side which are equal (i.e., $\mathbf{E_{ij}} = \mathbf{E_{ji}}$ since the matrix is symmetric). To reduce the memory requirements, a slight modification is introduced by Du et al. [9]. The following are defined, a public matrix $\mathbf{G}$ of size $(\lambda + 1) \times N$ and a private symmetric matrix $\mathbf{D}$ of size $(\lambda + 1) \times (\lambda + 1)$ where $\mathbf{D}$ entries are generated randomly. Also, $\mathbf{A} = (\mathbf{D} \cdot \mathbf{G})^\mathbf{T}$ of size $N \times (\lambda + 1)$ is defined. For a node $s_i$, row $\mathbf{R_i}$ in $\mathbf{A}$ and column $\mathbf{C_i}$ in $\mathbf{G}$ are selected. When two nodes $s_i, s_j$ would like eventually to communicate securely, they firstly exchange their $\mathbf{C_i}, \mathbf{Cj}$ then $k_{ij} = \mathbf{R_i} \cdot \mathbf{C_j}$ is computation in the side of $s_i$ and $k_{ji} = \mathbf{R_j} \cdot \mathbf{C_i}$ is computed in the side of $s_j$. Note that $k_{ji} = k_{ij}$ based on the symmetric property of $\mathbf{A}, \mathbf{D}, \mathbf{G}$. The second work by Blundo et. al. [3] proposed three protocols for secure dynamic conferences [3]. The 2-conferences protocol uses Symmetric Bivariate Polynomial (SBP) to distribute keys for $N$ nodes. The SBP has the following general form:

$$f(x,y) = \sum_{i,j=0}^{t} a_{ij} x^i y^j, \text{where } (a_{ij} = a_{ji}) \quad (1)$$

This polynomial is of degree $t$ where $t \leq N$. For a node $s_i$ with identifier $ID_i$, the share $g(y)$ expressed in Eqn 2 is calculated and loaded to $s_i$'s memory for generating future secret keys. Similarly, for two nodes $s_i, s_j$ that would like to communicate securely, $k_{ij} = g^i(j), k_{ji} = g^j(i)$ are evaluated locally in the corresponding sides and used respectively as secret keys.

$$g^i(y) = f(i,y) \quad (2)$$

In the sensor networks era, the early scheme of key pre-distribution specifically for WSN is introduced by ESCHENAUER and GLIGOR (a.k.a., EG scheme) [10]. In EG scheme, each node is let to randomly pick a key ring $S_k$ of size $k$ from a big keys pool of size $P$. The picking process maintains a probabilistic connectivity between any node and other nodes in the entire network. This connectivity is noted as $p_{actual}$ and defined as $p_{actual} = 1 - \frac{((P-k)!)^2}{(P-2k)!P!}$. If two nodes $s_i, s_j$ share a key $k : k \in S_{k_i} \cap S_{k_j}$ they both can use $k$ as a secret key. Otherwise, a path key establishment phase via single or several intermediate node(s) is performed. In [10] the usage of memory is reduced, however, a frail resiliency is resulted (i.e., if a small number of nodes are compromised, big communication fraction of non-compromised nodes is disclosed). To improve the resiliency, Chan et. al. proposed the Q-COMPOSITE scheme [4]. Using the same procedure of EG, a key between two nodes $s_i, s_j$ is available if and only if $S_{k_i} \cap S_{k_j}$ is a set of $q$ number of keys. If $\{k_1, \ldots, k_q\} \in \{S_{k_i} \cap S_{k_j}\}$, $\mathbf{hash}(\mathbf{k_1}||\mathbf{k_2}, \ldots, ||\mathbf{k_q})$ is used as $k_{ij}, k_{ji}$. Otherwise, intermediate node(s) are used. More analytical analysis on the probabilistic schemes is shown by KWANG and KIM in [13]

In addition to improving Blom's scheme in [2], Du et. al. proposed two schemes for key pre-distribution in [8, 9]. In the early one they introduced a deployment knowledge based scheme that improves BLOM's [2] by avoiding the unnecessary memory, communication, and computation with reasonable connectivity [8]. In [9], a multi-space matrix scheme based on [2, 10] is introduced. A $\tau$ number of private matrices $\mathbf{D}$ is selected randomly out of $\omega$ pre-constructed matrices providing connectivity $p_{actual}$ that is expressed as $p_{actual} = 1 - \frac{((\omega-\tau)!)^2}{(\omega-2\tau)!)\omega!}$. Different $\mathbf{A}$s' are created using the different $\mathbf{D}$s'. $\tau$ rows of the different $\mathbf{A}$s' are selected and assigned for each node. For $s_i, s_j$, if they have a common space $\tau_{i,j} : \tau_{i,j} \in \tau_i \cap \tau_j$, the rest of BLOM's scheme is performed. Otherwise, an intermediate node that has an intermediate space is used to construct a path key in a path key establishment phase. Even though much memory and communication are required and smaller connectivity is generated, this work provides a higher resiliency than in both of RG [4] and Chan et. al. [10]. For more accuracy, different deployment structures with practical error measurements and the probability distribution functions pdf based on [8] are introduced by Ito et. al. in [14]

Simultaneously, Liu et. al. proposed several schemes in [18, 19] for key distribution which are mainly based on Blundo et. al. [3]. In [18], the polynomial-based mechanism is used to assign several polynomials for each node in a similar way of EG scheme [10]. Two nodes can establish a secret key if and only if they share a common polynomial. Otherwise, the two nodes use an intermediate node for establishing a secure path.

The most significant work by Liu et. al. is in [18, 19]. In both works, for a network of size $N$, a two dimensional deployment structure that constructs a grid of $N^{1/2} \times N^{1/2}$ is suggested. Different nodes are deployed on different intersecting points and different polynomials are assigned for the different rows and columns of the grid. For two nodes $s_i$ and $s_j$, if $R_i = R_j$ or $C_i = C_j$, (i.e., both nodes have the same polynomial's share), a direct key establishment is performed. Else (i.e., $R_i \neq R_j$ and $C_i \neq C_j$), an intermediate node is used in an a path key establishment phase. In this work, even if a big fraction of nodes $p_c$ of the overall network size $N$ is compromised, the network remains connected via alternative intermediate nodes. The big fraction $p_c$ herein is measured to be $p_c \leq 60\%$ of $N$. Also, an n-dimensional scheme is introduced in [19]. Finally, the deployment knowledge for special purposes and applications using probabilistic manner has been studied in [26, 27] while a general security architecture has been proposed in [25].

## 1.2 Our Contributions

In this paper we introduce a new scheme using the *Hierarchical Grid* as a deployment framework and BLUNDO's scheme as key generator (a.k.a., keying material). Through this paper, our main contributions are the following:

- Provide a scalable, robust, and novel framework for the key pre-distribution that gives a perfect connectivity value (i.e., the connectivity is always equal to '1' using the single hop communication manner) to establish a pairwise key.

- Optimize the usage of the different network resources, mainly, communication overhead, memory usage, and required computation.

- Analyze and provide a mathematical model of our scheme's performance.

- Provide and discuss the alternative against any possible security attack against our scheme.

We take advantage of different flat deployment zones in a hierarchical grid representation for deploying the different sensor nodes. Based on each sensor node's location, several symmetric polynomials like these introduced in [3] are assigned to generate secret keys. Each polynomial in the assigned group for every sensor node is used for securing communication within a targeted zone. As a result, each node can communicate with any other node in the network using the shared keying material. We show how the connectivity approaches a prefect desirable level in both of the random and non-random cases. As well, we show the value of our scheme against some given attacks and study the its security under the amount of consumed resources.

## 1.3 Paper Structure

The rest of the paper is organized as follows. Section 2 introduces the notations and definitions which are used throughout the paper and Section 3 introduces our scheme. We consider an extensive analysis of our scheme's connectivity as a main interesting issue in Section 4. In Section 5, we consider the analysis of resources consumption and the security analysis in Section 6 considering several attacks. Finally, we introduce a comparison between our work and set of previous works in Section 7 followed by concluding remarks in Section 8.

## 2 Notations and Definitions

The following definitions and notations are used throughout the rest of this paper.

### 2.1 Definitions

**Definition 1** (Network order $n$). *a network design parameter that indicates the size of the network and the number of polynomials used in each sensor node.*

**Definition 2** (Basic grid or basic zone). *set of sensor nodes in a geographical area that initially use the same polynomial of degree $t_0$*

**Definition 3** (Polynomial Order $O$). *an integer that decides the scope where the polynomial is used to establish a secure pair-wise key, where $O \in \{1, 2, \ldots, n\}$. Each node has some minimum order of $1$ and maximum order of $n$*

**Definition 4** (Polynomial Degree $t_0$). *a security parameter that indicates the strength of the polynomial against the compromise and expresses how many different nodes that carry shares of this polynomial must be compromised for revealing the polynomial itself for an attacker. The subscription $0$ to $n$ expresses the order of the polynomial.*

### 2.2 Notations

The notation in table 1 is used through the rest of the paper.

## 3 HGBS for Pairwise KPD

Our scheme uses Blundo [3] as a keying material generating block to generate different secret keys for different nodes. The distribution of the keying material is performed on sensor nodes deployed in a *Hierarchical Grid* as shown in Figure 1. Our grid mainly considers the routing grid used in [17] with slight modification. This modification relies on using the duplication growth factor to move from an order to another. In our work, we aim to provide each sensor node with a set of different polynomials for establishing secret keys. The main rule of the different polynomials is to make several zones with varying number of nodes approachable by a given node. In the following subsections, we provide a description of our scheme including the following points: the deployment grid overview, node identification mechanism, keying material generation, secure key establishment and the scheme parameters adjustment.

**Table 1. Notation**

| Term | Indication |
|---|---|
| $n$ | the network order |
| $N$ | number of sensor nodes in the entire network |
| $m$ | number of sensor nodes in the basic grid $B_z$ |
| $k$ | modes distribution unit through the network |
| $B_z$ | basic zone (also, Basic Grid) |
| $O_x$ | order of the $x^{th}$ network grid |
| $t_0$ | degree of the basic polynomial in the $B_z$ |
| $t_n$ | degree of the polynomial for grid of order $n$ |
| $s_i, s_j$ | sensor nodes |
| $ID_i$ | identifier of the sensor node $i$ |
| $G_n$ | number of the Basic Zones in the network |

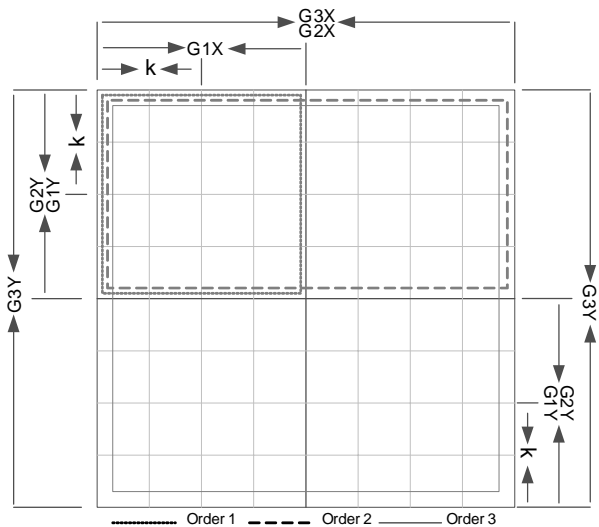

**Figure 1. Sensor nodes deployment in a hierarchical grids network**

### 3.1 Overview of the Deployment Grid

Consider a network that consists of $N$ sensor nodes. The different nodes are deployed in a network of grid structure as of Figure 1. In this deployment structure, the network is divided into $n$ hierarchical orders of grids. Each order $i$ consists of $2^{i-1}$ basic zone. The basic zone $B_z$ is a geographical region bounded by $[2k, 2k]$ dimensions (i.e., length and width). Also, $k$ is identified as the a uniform distribution unit of the sensor nodes in the WSN and the length unit as well. The number of the nodes $m$ in $B_z$ is $(2k)^2$. The order defined earlier is used to represent the growth of the network. The highest order $O_n$ contains $G_n = 2^{n-1}$ number of basic grids. The total number of nodes in the network is $N$ where $N = m \times G_n = (2k)^2 \times 2^{n-1}$. As shown in Figure 1, $B_z$ is any grid with the dimensions [G1X,G1Y] which has $O_1$. Similarly, the dimensions [G2X,G2Y] are considered for grids of order 2 ($O_2$), [G3X,G3Y] will be considered for $O_3$, and so on until $O_n$. An obvious note to mention here is that any zone which belongs to order $O_a$ includes twice as much as the number of nodes in zone $O_{a-1}$.

### 3.2 Node Identifier

Our scheme uses a smart identification material (ID) which is unique for each node through the network. The function of the ID in our scheme is to identify the node within the network, to represent the keying material (i.e. polynomials) of the node, and to provide "an extra sense" of the node location assuming a limited mobility.

The use of the hierarchical grid with a duplicating growth factor makes it possible to represent the different basic zones of Figure 1 in a binary tree as shown in Figure 2. In this tree, the height represents the maximum order and the number of leaves represents the number of basic zones in the network. For each leaf node, the attached numbers are sequences that represent the identifiers of different nodes within the same basic zones (i.e. local ID in a $B_z$ where $1 \leq ID_{local} \leq m$). The different polynomials are assigned to the internal nodes of the tree. In the tree, left branches are assigned to "0" bit value and right branches to "1" bit value. The final sensor node's ID is the binary string of tracing the path from the root to the end leaf that the sensor belongs to concatenated with the local ID. This structure of ID is shown in Figure 3. The length of this ID can be expressed as follows

$$|ID| = n + \lceil \lg(m) \rceil \qquad (3)$$

Where $m$ is is the number of nodes in the basic zone.

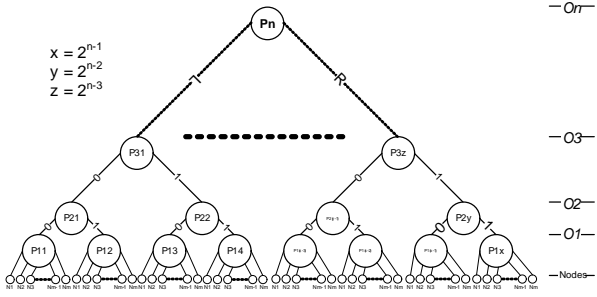

**Figure 2. Node ID generation determining node's location in WSN**

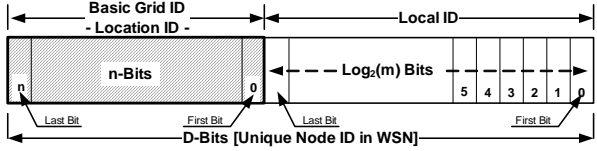

**Figure 3. Node ID structure in the hierarchical grid**

For a network with a large size, $n$ can be considered a constant for a flexible design that accepts dynamic network growth.

### 3.3 Key Material Assignment

Several keying material (or simply polynomials) is assigned for the different sensor nodes. Using the grid deployment structure as of Figure 1, different polynomials with the appropriate security parameter are assigned for the node based on which grids of which order it belongs to. Initially, SBP of degree $t_0$ which is assigned to the corresponding basic grid $B_z$ is assigned for establishing secret keys for the pairs of nodes within the same $B_z$. The usage of this polynomial will provide a value of $\frac{1}{2n-1}$ of direct connectivity. The other polynomials are used for the connectivity to reach the desirable one based on the node's location and granting the corresponding connectivity.

We assume that the nodes which are deployed within the same basic grid have the higher probability for communicating with each other and those outside the concerned basic grid have a less communication probability. This assumption is important since the usage of the polynomials with small degree that require a less computation power is so frequent and the usage of the higher degree polynomials is less frequent. The procedure of generating the keying material and the assignment for the different sensor nodes is shown in Algorithm 1.

---

**Algorithm 1**: Keying material assignment

**Input** : Network order $n$, set of all nodes' IDs, set of path IDs $d$, Network size $N$

**Output**: $n$ polynomials shares for each node $x$ as $k_x$.

1 **for** $i = 1$ **to** $n$ **do**
2     **for** $j = 0$ **to** $2^{i-1}$ **do**
3         $p[n-i+1][j] \Leftarrow$ SBP of degree $t$
4     **end**
5     **for** $x = 1$ **to** $N$ **do**
6         $k_x[n-i+1] = p[n-i+1][d/(2^{(i-1)}](ID_x, y)$
7     **end**
8 **end**

---

### 3.4 Key Establishment

To secure the communication between nodes $s_i, s_j$, secret key generation is required. Considering the ID of both nodes and the polynomial set generated by the algorithm in Algorithm 1, firstly, a polynomial $f^*(x, y)$ is selected out of the shared polynomials in the two nodes. The selected polynomial must be common in both nodes' polynomials with the minimum *t-degree* (i.e., referring to Figure 2, the the most close parent to the leaves of both nodes $s_i, s_j$). To establish the secure key, Algorithm 2 is applied. Note that, this algorithm is applied in both of $s_i, s_j$ to generate the pairwise key. Also, only the polynomial share is used after its evaluation in algorithm in Algorithm 1 as expressed in Eqn 2.

---

**Algorithm 2**: Key establishment procedure

**Input** : Path identifiers $d_i, d_j$, node's $s_j$ ID ($j$), set of node's $i$ polynomials shares; $k_i[]$

**Output**: Symmetric key $K_{ij}$.

1 Begin;
2 **for** $c = 0$ **to** $d_i.length - 1$ **do**
3     **if** $d_i[c] = d_j[c]$ **then**
4         $g(y) = k_i[c]$;
5         **Break**;
6     **end**
7 **end**
8 $k_{ij} = g(j)$

---

## 3.5 Parameters Adjustment

The critical parameter in our scheme that control the resources usage and resulting security is the polynomial degree $t_0$ and the relationship between $t_0$ and other polynomials' degrees in the different orders. A less important factor in our scheme's analysis is the communication traffic function (CTF).

The degree $t_0$ is totally dependent on the number of nodes in the basic grid [3]. In [29], the authors assigned $t_0$ to be 20. However, this assumption does not provide correlated dynamic security strength with the change of network size. Generally, if we consider $0 < \alpha \leq 1$ as a security parameter, $t_0$ can be expressed as $t = \alpha * m$ for more reliable security assumption. Using the same memory as in [12], $t_0$ can be assigned to $0.6 \times m$ which will make the basic zone secure till the compromise of $0.6m+1$ number of nodes that belong to the same grid. For a different numbers of nodes in the network, Figure 8 shows the required memory in KB to store the different polynomial coefficients. On the other hand, the remaining $(n-1)$-polynomials' degrees $t_1, t_2, \cdots, t_{n-1}$ are to follow one of the following approaches: (i) To have the value of $t_0$ and the growth of the network order will lead to the same value of the polynomial growth. (ii) To consider the different $t$ degrees independently.

For the communication traffic function (CTF), our deployment scenario considers that the nodes which are mostly to communicate are the neighbors in the same basic grid while other nodes in other grids have a less traffic fraction. In the analysis, we consider two different functions: The geometric series distribution function and the exponential distribution functions. Note that, the summation of probabilities for the communication from a given zone that represents a set of given nodes to all other zones is equal to one (i.e. $\sum_{i=1}^{n} p_{pdf} = 1$). For a geometric CTF, $c$ is determined satisfying Eqn 4.

$$\text{CTF} = \sum_{i=1}^{n} \left(\frac{c}{2^{i-1}}\right) = 1 \qquad (4)$$

## 4 Connectivity

Dividing the network hierarchically provides a connectivity using more than one keying material (i.e., symmetric polynomials). This connectivity enables different nodes belonging to different basic grids to communicate securely in 1-hop fashion. Let us consider $C$ as the provided connectivity. Based on the structure of the grid, the polynomial assigned to $B_z$ provides connectivity of $C_1$ which is expressed as $\frac{m}{m \times 2^{n-1}} = \frac{1}{2^{n-1}}$. In general terms, the polynomial for the $i^{th}$ order grid provides connectivity of $\frac{m \times 2^{i-1}}{m \times 2^{n-1}} = \frac{1}{2^{n-i}}$ for the nodes that belong to it. Thus, for the highest order, the provided connectivity is $\frac{m \times 2^{n-1}}{m \times 2^{n-1}} = 1$ which exclusively includes all of the below orders' connectivity (i.e., $C_i$ for $i = 0$ to $n-1$).

For a more general conception, we consider two types of connectivity. The first type conceptualize the connectivity that considers sensor nodes $s_i, s_j$ in the network regardless to their location while the second type consider sensor nodes according to their precise minimum-weighted polynomial that they share. For the first case, nodes are randomly deployed in the field and the traffic does not have a regular pattern. The connectivity is expressed as the probability $p_z$ for the two nodes $s_i, s_j$ to be in the same grid. Similarly, this can be formulated and expressed as the probability for two nodes within some order order to exactly belong to the same minimum order $z$ which enables establishing keys with minimal weighted polynomial. Also, it can be expressed as the probability of randomly picking two nodes with the condition that they belong to the same order. This probability is shown in Figure 6

$$p_z = \binom{2^{(z-1)}m}{2}\binom{N}{2}^{-1} = \binom{2^{(z-1)}m}{2}\binom{2^{(n-1)}m}{2}^{-1}$$
$$= \frac{(2^{z-1}m)(2^{z-1}m-1)}{(2^{n-1}m)(2^{n-1}m-1)} = \left(2^{z-n}\right)\left(\frac{2^{z-1}m-1}{2^{n-1}m-1}\right) \qquad (5)$$

Consider $a, b$ as two integers such that $a \geq b > 1$, $\frac{a}{b}$ is always *greater than or equal to* $\frac{a-1}{b-1}$. Applying this to Eqn 5 we get that $\frac{(2^{z-1}m-1)}{(2^{n-1}m-1)} \leq 2^{z-n}$. From all, we get:

$$p_z \leq \left(2^{(z-n)}\right)^2 \qquad (6)$$

On the other hand, the second type of connectivity considers two nodes $s_i, s_j$, where their used polynomial to generate secret key is previously determined based on the fixed deployment structure and traffic model. A connectivity $C_i$ is determined as the provided certain connectivity provided by the minimal network order that the two nodes belong to. This is typically equivalent to the connectivity provided by any polynomial assigned for any the given polynomial order as in Eqn 7.

$$C_i = \frac{2^{i-1}}{2^{n-1}} = 2^{i-n} \qquad (7)$$

For both $p_z$ and $C_i$, the overall connectivity is defined as the provided ability to each node to communicate securely with other nodes in secure manner regardless to the order they belong to and the shared polynomial they use. In other words, the overall connectivity is determined as the value of $p_z$ or $C_i$ when $z$ or $i$ go to $n$ as follows:

$$C_n = \left(2^{(z-n)}\right)^2 \bigg|_{z=n} = \left(2^{(i-n)}\right)\bigg|_{i=n} = 1 \qquad (8)$$

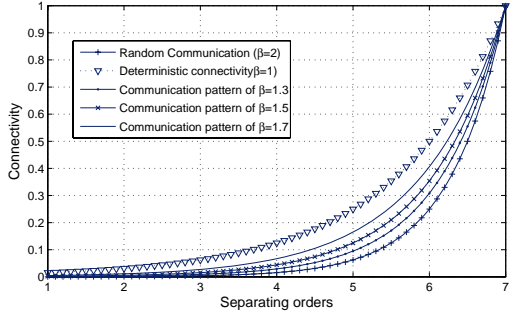

**Figure 4. Connectivity with different traffic parameters**

For a general connectivity form that considers degree of randomness which determines each node's knowledge of the deployment structure given and the accuracy of communication model description [22], we define $\beta$ as communication pattern's parameter where $C = (2^{i-n})^\beta$ and $1 \leq \beta \leq 2$. Figure 4 shows the connectivity provided by the order's polynomials according to different $\beta$ values.

## 5 Overhead Analysis

In this part we consider our scheme's overhead. This overhead is mainly represented by the memory required for the keying material representation, the computation required for a single bivariate polynomial's evaluation on $\mathbf{GF}(q)$ in a single variable, and the communication required for exchanging the concerned sensor node's identifiers.

### 5.1 Network Capacity

Our scheme uses the different resources of the network in a reasonable manner. The reduction in using any resource can affect other correlated resources and downgrade the overall performance. In this section, we measure the cost of our scheme by analytical and mathematical formulas in terms of the network resources. From the details above, the total network capacity $N$ can be expressed as

$$N = 2^{(n-1)} \times (2k)^2 \qquad (9)$$

Where $n$ is the largest polynomial order in the network and $k$ is the distribution unit of nodes. The relationship between $k$ and $n$ for different network size $N$ and ranging $n$ is shown in Figure 5 and Figure 6.

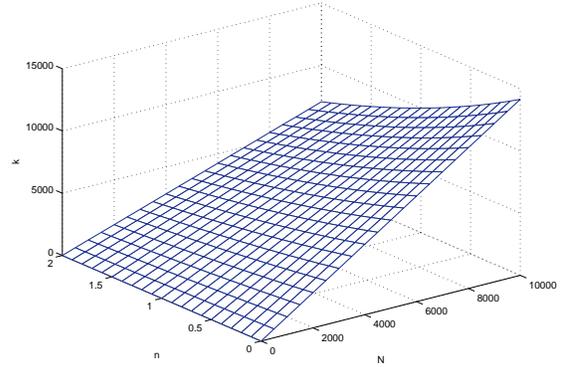

**Figure 5. The relationship between $n, k, N$ for $0 < n \leq 2$.**

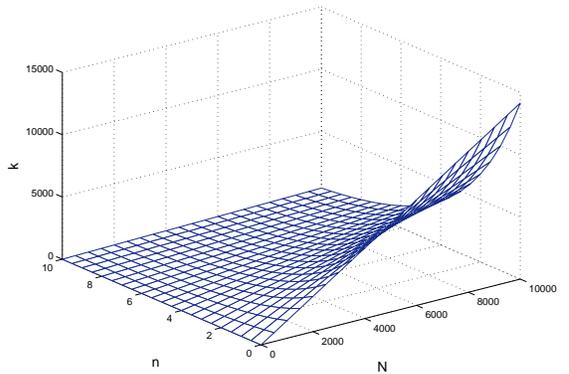

**Figure 6. The relationship between $n, k, N$ for $0 < n \leq 10$.**

### 5.2 Memory Overhead

The amount of memory which is required per node is to represent the node ID shown in Eqn 3 and the different n-polynomials' shares. For a polynomial $f(x,y)$ of degree $t_0$ whose coefficients in $\mathbf{GF}(q)$, $(t_0 + 1) \times \lg(q)$ bits are required as a representation space. For the memory use, we introduce two different approaches:

1. To assign different degrees for the different polynomials regardless to the number of the distributed shares of the polynomial.

2. To make the growth of the different polynomials' degrees same as that of the growth of the number of the nodes that hold those polynomials' shares. Thus, all polynomials in the first order have degree $t_0$ and the $i^{th}$ order polynomials have degree of $2^{i-1} \times t_0$.

The first case cost in bits is represented in Eqn 10 as follows:

$$M_1 = n + \left\lceil \lg\left(\frac{N}{2^{n-1}}\right)\right\rceil + n\left(\frac{\alpha N}{2^{n-1}} + 1\right)\lg(q) \quad (10)$$

The first two terms are for the ID representation and the third term for $n$-polynomials representation. The second case is shown in Eqn 11 where the third term is the summation of the required memory to represent $n$ polynomials of different degrees and $\alpha$ is the security parameters, $P_{weight}$ is $f(x, ID)$'s $(t_0 + 1)$-coefficients representation memory.

$$M_2 = n + \left\lceil \lg\left(\frac{N}{2^{n-1}}\right)\right\rceil + P_{weight}\sum_{i=1}^{n}(2^{i-1})$$
$$= n + \left\lceil \lg\left(\frac{N}{2^{n-1}}\right)\right\rceil + (2^n - 1)\left(\frac{\alpha N}{2^{n-1}} + 1\right)\lg(q) \quad (11)$$

Also, the memory requirement in Eqn 11 can be considered as a geometric series with base $r$ in terms of the highest order's polynomial (i.e. $r = \frac{1}{2}$). For this representation, the summation is held for $n$ terms giving the following:

$$M_3 = \left(1 + \frac{1}{2} + \cdots + \frac{1}{2^{n-1}}\right)(\alpha N \times \lg(q))$$
$$= (\alpha N \times \lg(q)) \times \left[\sum_{i=0}^{n-1}\left(\frac{1}{2}\right)^i\right] \quad (12)$$

Recall that $r : -1 < r < 1$, the summation of the first $n$ terms is $S_n = \left(\frac{1-r^{n+1}}{1-r}\right)$. This gives the following:

$$M_3 = (\alpha N \lg(q))\left(\frac{1 - \left(\frac{1}{2}\right)^{n+1}}{1 - \frac{1}{2}}\right) \quad (13)$$

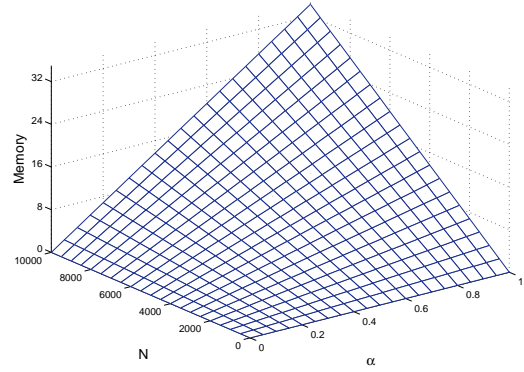

**Figure 7. The required memory for different security parameters and network size in our scheme.**

Figure 7 shows the required memory for different network size $N$ and security parameter $\alpha$. Also, Figure 8 shows the required memory for the Blundo's single polynomial representation. Note that, the memory requirements is linearly dependent on both $N$ and $\alpha$.

### 5.3 Computation Overhead

Each time a key is required, the evaluation of polynomial $f(x, ID)$ of degree $t$ is performed in single variable. Due to the difference of $t$ degree shown earlier, different scenarios are considered. In case of using the first memory scenario, the required computation can be summarized in a single polynomial $f(x, ID)$ of degree $t_0$ evaluation regardless to the degree. In the second scenario where we assign different degree according to the growth factor for the different network orders, the communication pattern and communication probability function based on the location determines the required polynomial to be computed and the required computational power.

From other point, evaluating a polynomial of degree $t$ requires $2t$ number of multiplications [1]. Based on [16]; using the long integer multiplication, it requires 64 number of 8-bit word multiplication to multiply two integers in finite field of $q = 2^{64}$. In contrast, it takes only 27 word multiplication on the same platform using Karatsuba-Ofman algorithm [16]. Thus, the required word multiplication for multiplying two integers of length 16 and 64 bits consumes 16 8-bit word multiplications. This type of different length operations is required for accumulating a long enough key that fits with the required standard key length (e.g. DES of 56 bit, AES of 128 and RC5 of 64)[2].

---
[1] Recall that the computational weight of $n$-bit $n$-integers addition is equivalent to two integers multiplication with the $n$ length. Initially, the required number of multiplications is $2t - 1$ and the additional one due to the addition operation [16]

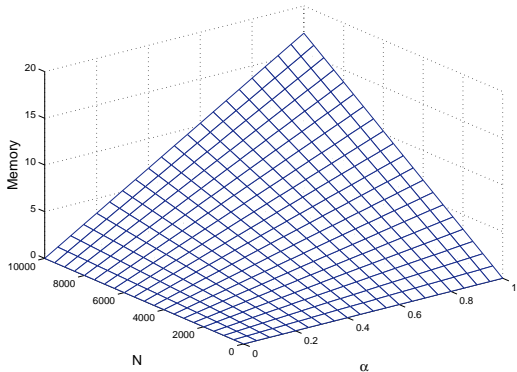

**Figure 8. Memory requirements for storing the polynomial's coefficients in Blundo's scheme with varying $\alpha$.**

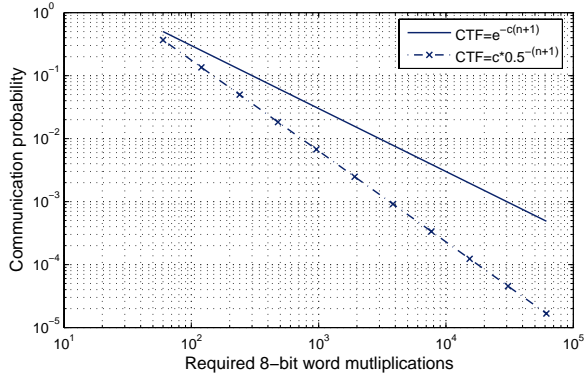

**Figure 9. Computation overhead for two different Communication Traffic Functions. The used parameters are $t_0 = 60$, $\beta = 0.6$, $n = 11$ which supports up to $N = 204800$.**

$$CP_{avg} = \left(\sum_{i=1}^{n}(p_i CP_{t_i})\right) + c \quad (14)$$

Eqn. 14 expresses the required computation in terms of the number of multiplications in $\mathbf{GF}(q)$, where $c$ is computational power required for two binary strings comparison. These strings represent the polynomial path identifier part. Also, $p_i$ is the probability that two nodes reside in different $(i-1)^{th}$ grids and $CP_{t_i}$ is the required computational power for the $i^{th}$ order polynomial evaluation. For two different communication traffic functions, Figure 9 shows the computational consumption growth curve in terms of the number of multiplications in $\mathbf{GF}(q)$ according to the growth of the network size.

### 5.4 Communication Overhead

Our scheme does not require any extra means of communication except of the nodes' identifiers exchange which costs $\lg(N)$ bits transmission as discussed early in the ID representation requirement. The ID itself is expressed in a general form as of Eqn 3.

---

[2]The RC5 [31] is the most likely to be used on the typical sensor node platform because of its small code size without a need for extra tables [25]. Thus, the largest expected finite field is of 64 bits length.

### 5.5 Reducible Memory Schemes

Since the memory limitation is the bottleneck for any successfully design in wireless sensor network, any set of polynomials with an order greater than a desirable $d$ such that $1 < d \leq n$ can be discarded by giving up the direct connectivity. To ensure an alternative connectivity through intermediate node, a virtual grid as of [18] can be applied. Once this kind of grid is applied, the higher order polynomials are removed and replaced by smaller polynomials that represent the virtual grid. Therefore, the final required memory is dependent upon how many orders are discarded but always less than the required memory overhead expressed in 12 or 13.

## 6 Security Analysis

The security of our scheme follows the same analysis introduced in [3, 18]. The security of all of these schemes is based on that a network that uses a bivariate polynomial of degree $t$ is secure against the compromise as long as the number of revealed shares is less than or $t + 1$. In this section, we consider several attacking scenarios against our scheme. This includes the node replication attack [24], the Sybil attack [23], Denial of Messages (DoM) Attacks [21], and Denial of Service (DoS) Attacks [34].

## 6.1 Compromising Effects and Resiliency

### 6.1.1 An Attack Against $N_c$ Number of Nodes

In case of compromising a set of nodes whose size is $N_c$ that is less than $t_0$, the fraction of the affected nodes other than those which are compromised is 0. This applies for any kind of attack strategy including the random and selective one [12].

### 6.1.2 An attack against $B_z$

An attack against the single basic zone can be successfully performed through the compromise of number of nodes $N_c$ where $N_c > t + 1$. By compromising $t + 1$ number of sensor nodes and revealing their secret shares, the main polynomial of the concerned node can be recovered[30]. Even though, this attack looks more difficult due to that $2^{n-1}$ number of polynomials of degree $t$ within the network where the probability $p_r$ for $t_0$ nodes to be belonging to the same polynomial shares is in Eqn 15.

$$p_r = 1 - \sum_{i=0}^{t_0-1} \binom{N_c}{i} \left(\frac{m}{N}\right)^i \left(\frac{N-m}{N}\right)^{N_c-i} \quad (15)$$

The probability $p_r$ indicates that the number of sensor nodes to be compromised by revealing its own secret shares should be big enough to guarantee that the at least $t_0$ number of shares belong to a specific polynomial. This primarily based on the polynomial degree $t_0$ and $\alpha$, however, over 60% of the network size is a reasonable threshold for $\alpha = 0.6$

### 6.1.3 An attack against the whole network

The attack against the whole network can not be in synchronized way. However; in the worst case, it is possible to compromise the entire network by compromising all of the polynomials $f(x, y)$ of $t_0$ one by one. To compromise $B_z$ requires $t_0$ nodes to be compromised. Since the network consist of $G_n$ different $B_z$, it requires to compromise $G_n \times t_0$ which is a big fraction (i.e. more than 60% of the network size). Without this amount, the fraction of affected nodes will be less than 50% of the network size.

### 6.1.4 Selective versus random node attack

Even if the nodes are deployed randomly, the knowledge of the nodes deployment and the assigned polynomials for each group and the ability to distinguish the different nodes based on their $B_z$ enables a selective attack that ease the attackers task. On the other hand, the random attack where the attacker's knowledge about the deployment strategy of the several nodes makes it harder to reveal a given polynomial that generates secure keys for a given polynomial as shown previously in Eqn 15.

### 6.1.5 Sybil and node replication attacks

There are two problems belonging to the dynamic growth of sensor network. Sybil attack [23] is done fallaciously by using more than one ID for the same node $j$ and node replication attack [24] which is performed using the same ID more than one time in the network. Our work resists against these threats because it requires a structured ID which is unique with a uniform structure over the entire network. When an attacker fabricates a structured ID, it should follow the limited structure shown previously and deploy the node in specific area to communicate within the same $B_z$.

### 6.1.6 DoM and DoS attacks

Denial of Messages [21] is the ability of some nodes (i.e., attacker's nodes) to deprive others of receiving some broadcast messages. Our framework does not require any broadcast capability. If any, it will be mainly used within the same grid. Thus, the DoM attack will only affect a small fraction of the whole network. An example of the Denial of Service[34] is "attempts to prevent a particular individual from accessing a service" and this mainly happens due to a heavy communication or computation because of the key generation or any outsider reason like attacker messages flooding. In our scheme, all of the computation and communication operations are small, and take short time. In the second case, to perform a DoS, node replication attack is required.

### 6.1.7 Man In the Middle Attack (MITM)

Under the assumption that the radio coverage is enough to enable the usage of the different polynomials for different targets, the man in the middle attack is impossible. In the case of the reduced memory schemes, the man in the middle attack is possible with small probability due to that the intermediate nodes are used for a limited communication fraction of forwarding. As well, each attacker who would like to deploy his own sensor nodes to perform the MITM should know the geographical location where to deploy the different sensor nodes.

**Table 2.** Comparison between our scheme and a set of other schemes in terms of the resources usage and the resulting connectivity. The connectivity for the probabilistic key pre-distribution schemes is probabilistic while it is certain for other schemes including ours. Also, the polynomial degree $t$ differs as shown earlier based on $\alpha$.

| Scheme | Communication | Computation | Memory | Connectivity |
|---|---|---|---|---|
| GBS [18] | c | SBP Evaluation | ID+2 SBP | $\frac{2}{N^{1/2}-1}$ |
| 3D-GBS [18] | c | SBP Evaluation | ID+3 SBP | $\frac{3}{N^{2/3}+N^{1/3}+1}$ |
| Plat-Based [22] | c | SBP Evaluation | ID+3 SBP | $\frac{3}{N^{1/3}}$ |
| EG [10] | $c\log_2(S_k)$ | $\frac{(2c+p-p_k)}{2}\log_2(c)$ | $S_k$ keys | $1 - \frac{((P-k)!)^2}{(P-2k)!P!}$ |
| CPS [4] | c | c | $S_k$ keys | $\frac{m}{N}$ |
| DDHV [9] | $c\log_2(n \times \tau)$ | 2 vectors mult. | $\tau + 1$ vectors | $1 - \frac{((\omega-\tau)!)^2}{(\omega-2\tau)!\omega!}$ |
| HGBS | c | SBP Evaluation | ID+$n$ SBP | 1 |

## 6.2 Blocked Traffic and Its Recovery

This paper mainly introduces a new framework for the key pre-distribution, deployment and smart location based identification. However, when we applied BLUNDO's scheme [3], we obtained that even though the $i^{th}$ order polynomial where $1 < i \leq n$ is compromised, this will not affect the other network except of that amount of traffic (links) within the $i^{th}$ order grid. Assume the $i^{th}$ order SBP is compromised, the fraction of the blocked traffic will be $\frac{m \times 2^{i-1}}{m \times 2^{n-1}} \times p_i = \frac{1}{2^{n-i}} \times p_i$ where $p_i$ is the fraction of traffic between nodes resides in different the $(i-1)^{th}$ order grids. Using the current $p_i = \frac{1}{2^{i-1}}$ distribution will guarantee that the blocked communication is always constant value regardless to $i$ value.

On the recovery; when $t+1$ nodes are compromised, an alternative secure SBP can be used. In the case that an SBP of the $c^{th}$ order grid is compromised, the SBP for the $(c+1)^{th}$ order grid is used till the system recovery and assigning another polynomial to the affected grid. In case of the highest order's polynomial compromise, the amount of traffic compromised will be only $\frac{1}{2}p_{(i=n)}$. If we assume that the fraction is decreased by half whenever the order of grid increases by 1, the amount will be will be $\frac{1}{2^{n-1}}$ for $p_n$. However, the internal network connectivity will not be affected. Moreover, the majority of the secure traffic in the network will not be broken since the deployment framework guarantees that most of the traffic is in $B_z$.

## 7 Comparison With Others

We selected GBS [18], Multi-space [9], EG [10], Q-Composite and RPS [4] for the comparison with our scheme. The compared features are communication, computation and memory. Table 2 shows this comparison in terms of those resources. In our scheme, memory, computation and communication requirements are shown in Eqn 10 to Eqn 14

**Remark:** the constant value of communication in GBS depends on whether it's possible to construct a direct key or not. In case of using an intermediate node, the communication cost of the intermediate should be considered. The amount of communication traffic in our scheme is always constant because of its nature.

## 8 Conclusion

We proposed a novel framework for the secure key pre-distribution in the WSN. Our proposed scheme uses a hierarchical grid for the sensor nodes deployment that bounds the heavily communicated nodes in one basic grid that has strong secure keying material.

We also designed an ID structure which is unique for the node and expresses the location as well as the keying material to be used. To measure the performance of our framework, we used BLUNDO's [3] as a keying material generator block. Mathematical analysis of the computation, communication and memory was provided. The different possible attacks were lightly touched. The performance shown comparison expressed the value of our framework.

## References

[1] I. Akyildiz, W. Su, Y. Sankarasubramaniam, and E. Cayirci. A survey on sensor networks, 2002.

[2] R Blom. An optimal class of symmetric key generation systems. In *Proc. of the EUROCRYPT*

*84 workshop on Advances in cryptology: theory and application of cryptographic techniques*, pages 335–338, New York, NY, USA, 1985. Springer-Verlag New York, Inc.

[3] Carlo Blundo, Alfredo De Santis, Amir Herzberg, Shay Kutten, Ugo Vaccaro, and Moti Yung. Perfectly-secure key distribution for dynamic conferences. In *CRYPTO*, pages 471–486, 1992.

[4] Haowen Chan, Adrian Perrig, and Dawn Xiaodong Song. Random key predistribution schemes for sensor networks. In *IEEE Symposium on Security and Privacy*, pages 197–, 2003.

[5] David Culler, Deborah Estrin, and Mani B. Srivastava. Overview of sensor networks. In *IEEE Computer Society*, pages 41–49, 2004.

[6] Jing Deng, Richard Han, and Shivakant Mishra. Defending against path-based dos attacks in wireless sensor networks. In *SASN*, pages 89–96, 2005.

[7] Whitfield Diffie and Martin E. Hellman. New directions in cryptography. *IEEE Transactions on Information Theory*, IT-22(6):644–654, 1976.

[8] Wenliang Du, Jing Deng, Yunghsiang S. Han, Shigang Chen, and Pramod K. Varshney. A key management scheme for wireless sensor networks using deployment knowledge. In *INFOCOM*, 2004.

[9] Wenliang Du, Jing Deng, Yunghsiang S. Han, Pramod K. Varshney, Jonathan Katz, and Aram Khalili. A pairwise key predistribution scheme for wireless sensor networks. *ACM Trans. Inf. Syst. Secur.*, 8(2):228–258, 2005.

[10] Laurent Eschenauer and Virgil D. Gligor. A key-management scheme for distributed sensor networks. In *ACM CCS*, pages 41–47, 2002.

[11] Nils Gura, Arun Patel, Arvinderpal Wander, Hans Eberle, and Sheueling Chang Shantz. Comparing elliptic curve cryptography and rsa on 8-bit cpus. In *CHES*, pages 119–132, 2004.

[12] Dijiang Huang, Manish Mehta, Deep Medhi, and Lein Harn. Location-aware key management scheme for wireless sensor networks. In *SASN*, pages 29–42, 2004.

[13] Joengmin Hwang and Yongdae Kim. Revisiting random key pre-distribution schemes for wireless sensor networks. In *SASN*, pages 43–52, 2004.

[14] Takashi Ito, Hidenori Ohta, Nori Matsuda, and Takeshi Yoneda. A key pre-distribution scheme for secure sensor networks using probability density function of node deployment. In *SASN*, pages 69–75, 2005.

[15] Holger Karl and Andreas Willig. *Protocols and Architectures for Wireless Sensor Networks*. John Wiley & Sons Ltd., ISBN-13 988-0-470-09510-2, 2005.

[16] Donald Knuth. *The Art of Computer Programming: Semi-numerical Algorithms, volume Vol.2, third edition*. Addison-Wesley, ISBN: 0-201-89684-2, 1997.

[17] Jinyang Li, John Jannotti, Douglas S. J. De Couto, David R. Karger, and Robert Morris. A scalable location service for geographic ad hoc routing. In *MOBICOM*, pages 120–130, 2000.

[18] Donggang Liu and Peng Ning. Establishing pairwise keys in distributed sensor networks. In *ACM CCS*, pages 52–61, 2003.

[19] Donggang Liu, Peng Ning, and Rongfang Li. Establishing pairwise keys in distributed sensor networks. *ACM Trans. Inf. Syst. Secur.*, 8(1):41–77, 2005.

[20] David J. Malan, Matt Welsh, and Michael D. Smith. A public-key infrastructure for key distribution in tinyos based on elliptic curve cryptography. In *First IEEE Int. Conf. on Sensor and Ad Hoc Comm. and Networks*, pages 71–80, 2004.

[21] Jonathan M. McCune, Elaine Shi, Adrian Perrig, and Michael K. Reiter. Detection of denial-of-message attacks on sensor network broadcasts. In *IEEE Symposium on Security and Privacy*, pages 64–78, 2005.

[22] Abedelaziz Mohaisen, YoungJae Maeng, and DaeHun Nyang. On the grid based key pre-distribution: Toward a better connectivity in wireless sensor networks. In *SSDU 2007, to appear*, pages 00–00, 2007.

[23] James Newsome, Elaine Shi, Dawn Xiaodong Song, and Adrian Perrig. The sybil attack in sensor networks: analysis & defenses. In *IPSN*, pages 259–268, 2004.

[24] Bryan Parno, Adrian Perrig, and Virgil D. Gligor. Distributed detection of node replication attacks in sensor networks. In *IEEE Symposium on Security and Privacy*, pages 49–63, 2005.

[25] Adrian Perrig, Robert Szewczyk, J. D. Tygar, Victor Wen, and David E. Culler. Spins: Security protocols for sensor networks. *Wireless Networks*, 8(5):521–534, 2002.

[26] Roberto Di Pietro, Luigi V. Mancini, and Alessandro Mei. Random key-assignment for secure wireless sensor networks. In *SASN*, pages 62–71, 2003.

[27] Roberto Di Pietro, Luigi V. Mancini, and Alessandro Mei. Efficient and resilient key discovery based on pseudo-random key pre-deployment. In *IPDPS*, 2004.

[28] Ronald L. Rivest, Adi Shamir, and Leonard M. Adleman. A method for obtaining digital signatures and public-key crypto-systems. *CACM*, 26(1):96–99, 1983.

[29] Stefan Schmidt, Holger Krahn, Stefan Fischer, and Dietmar Wätjen. A security architecture for mobile wireless sensor networks. In *ESAS*, pages 166–177, 2004.

[30] Adi Shamir. How to share a secret. *Commun. ACM*, 22(11):612–613, 1979.

[31] William Stallings. *Cryptography and Network Security: Principles and Practice, 3rd edition*. Prentice Hall, ISBN: 0138690170, 2003.

[32] Arvinderpal Wander, Nils Gura, Hans Eberle, Vipul Gupta, and Sheueling Chang Shantz. Energy analysis of public-key cryptography for wireless sensor networks. In *PerCom*, pages 324–328, 2005.

[33] Ronald J. Watro, Derrick Kong, Sue fen Cuti, Charles Gardiner, Charles Lynn, and Peter Kruus. Tinypk: securing sensor networks with public key technology. In *SASN*, pages 59–64, 2004.

[34] Anthony D. Wood and John A. Stankovic. Denial of service in sensor networks. *IEEE Computer*, 35(10):54–62, 2002.